\documentclass[article,prl,balancelastpage,twocolumn,showpacs]{revtex4}
\usepackage{amssymb}
\usepackage{dcolumn}
\usepackage{graphicx}
\usepackage{acronym}

\input{tcilatex}

\begin{document}

\title{ Quantum states in rotating electromagnetic fields}
\author{B. V. Gisin }
\affiliation{IPO, Ha-Tannaim St. 9, Tel-Aviv 69209, Israel. E-mail:
borisg2011@bezeqint.net}
\date{\today }

\begin{abstract}
\noindent We describe a new class of exact square integrable solutions of
the Pauli and Dirac equation in rotating electromagnetic fields. Solutions
obtained by putting equations in the stationary form with help of a
coordinate transformation corresponding to the transition into a rotating
frame. The transformation is assumed to be Galilean one however a
non-Galilean transformation is of particular interest for such solutions.
Obtained solutions, especially of Dirac's equation, are valid for arbitrary
values of parameters and may be tested experimentally.\hfill
\end{abstract}

\pacs{03.65.Pm, 03.65.Ta, 31.30.jx, 06.20.Jr\hspace{20cm}}
\maketitle

\section{Introduction}

Exact solutions of field equations are of fundamental importance as in
theoretical as experimental physics. Such solutions of the Pauli and Dirac
equation are described in a variety of books (see, for example, \cite{Lan}).
However, as far as the author knows, there are no publications on
two-dimensional (2D) square integrable solutions of these equations in
rotating electromagnetic fields with a constant magnetic field along the
rotation axis. In this paper we present such solutions for the Pauli
equation in a rotating magnetic field and for the Dirac equation in the
field of a traveling circularly polarized electromagnetic wave.

For finding the solution the translation of spinor into a rotating frame is
necessary. This translation must be accompanied by the transformation of
coordinates and time. These manipulations turn the non-stationary into
stationary problem. However what must be the transformation? In the appendix
we discuss some physical arguments for the choice of the reasonable
transformation.

The Galilean or Lorentz transformation is the term pertaining to mechanics.
We use here this term keeping in mind that the role of velocity is played by
frequency.

\section{The Pauli equation}

We start from the Pauli equation%
\begin{equation}
i\hbar \frac{\partial }{\partial t}\Psi =\frac{1}{2m}(\mathbf{p}-\frac{e}{c}%
\mathbf{A})^{2}\Psi -\mu (\mathbf{\sigma H)}\Psi  \label{Paul}
\end{equation}%
in standard notations. $\mathbf{A}$ is the potential: $A_{x}=-H_{z}y/2,$ \ $%
A_{y}=H_{z}x/2,$ $A_{z}=H[-x\sin (\Omega t)+y\cos (\Omega t)]$, $H_{z}$ is
the constant component of the magnetic field along the $z$-axis, $H$ is the
amplitude of the rotating transverse component$,$\emph{\ }$\hbar ,m,e,c,\mu $%
\emph{\ }\ and $\sigma _{k}$ is the Plank constant, mass, charge, speed of
light, magnetic moment and Pauli's matrix respectively, $\Omega $ is the
frequency of rotation. $\Omega $ may take as positive as negative values.
Negative $\Omega $ corresponds to the reverse rotation. We consider
solutions with the definite momentum $p$ along the $z$-axis. Solutions must
be continuous and square integrable over all the cross-section transversely
to this axis.The potential correspond to the magnetic field%
\[
H_{z}=A_{y,x}-A_{x,y}=H_{z},\ H_{y}=H\sin (\Omega t),\ H_{x}=H\cos (\Omega
t), 
\]

The search for solutions consists of the wave function translation into the
rotating frame $\tilde{\Psi}=\exp (i\sigma _{3}\Omega t/2)\Psi $and a
transformation of coordinates%
\begin{equation}
\tilde{\varphi}=\varphi -\Omega t,\text{ }\tilde{r}=r,\text{ }\tilde{z}=z,%
\text{\ }\tilde{t}=t,  \label{D2}
\end{equation}%
where\emph{\ }$r^{2}=x^{2}+y^{2},$\emph{\ }$\tan \varphi =y/x$\emph{. }Above
two operations reduce the Pauli equation to the stationary case. Further
this equation is putted to the diagonal form $\tilde{\Psi}_{d}=\exp (i\sigma
_{2}\gamma /2)\tilde{\Psi},$ where \ 
\begin{equation}
\tan \gamma =\frac{2m}{\hbar ^{2}}\frac{\mu H}{\Delta },\text{ \ }\Delta =%
\frac{m\Omega }{\hbar }+\frac{2m}{\hbar ^{2}}\mu H_{z}.  \label{diag}
\end{equation}%
After that we use new coordinates 
\begin{equation}
\tilde{x}=r\cos \tilde{\varphi},\ \tilde{y}=r\sin \tilde{\varphi},
\label{cc}
\end{equation}%
and investigate stationary states $\tilde{\Psi}_{d}=\exp (-iE\tilde{t}/\hbar
+ipz/\hbar )\psi $, where $E$ is the "energy" in the rotating frame. Finally
denote 
\begin{eqnarray}
g_{1} &=&\frac{e^{2}}{\hbar ^{2}c^{2}}\frac{1}{4}H_{z}^{2},\text{ \ }g_{2}=%
\frac{e^{2}}{\hbar ^{2}c^{2}}(\frac{1}{4}H_{z}^{2}+H^{2}),  \label{ds1} \\
\text{ }b &=&\frac{2pe}{\hbar ^{2}c}H,\text{ \ }\epsilon =\frac{2m}{\hbar
^{2}}E-\frac{p^{2}}{\hbar ^{2}}+\sigma _{3}\rho ,  \label{ds2} \\
\text{\ \ }f &=&\frac{2m}{\hbar }\Omega +\frac{eH_{z}}{\hbar c},\text{ \ }%
\rho =\sqrt{(\frac{2m}{\hbar ^{2}}\mu H)^{2}+\Delta ^{2}}.  \label{ds3}
\end{eqnarray}%
Then the Pauli equation may be written in the following stationary form%
\begin{equation}
\psi _{,\tilde{x}\tilde{x}}+\psi _{,\tilde{y}\tilde{y}}-if\left( \tilde{x}%
\psi _{,\tilde{y}}-\tilde{y}\psi _{,\tilde{x}}\right) -(g_{1}\tilde{x}%
^{2}+g_{2}\tilde{y}^{2}-b\tilde{y}-\epsilon )\psi =0.  \label{eq0}
\end{equation}%
In this equation the comma means the differentiation. In Eq. (\ref{eq0}) two
components of the wave function are defined by independent equations. Exact
solutions for every component of the wave function exist in the form $\psi
^{\pm }=\psi _{n}^{\pm }\exp D,$ where $\pm $ corresponds to the sign of
diagonal components of $\sigma _{3}$, 
\begin{equation}
D=\frac{1}{2}d_{11}\tilde{x}^{2}+d_{12}\tilde{x}\tilde{y}+\frac{1}{2}d_{22}%
\tilde{y}^{2}+d_{1}\tilde{x}+d_{2}\tilde{y},  \label{sform}
\end{equation}%
$\psi _{n}^{\pm }$ is a polynomial in $\tilde{x}$ and $\tilde{y}$, the
quantum number $n\geq 0$ is defined as maximal number $n=n_{1}+n_{2}$ in the
product $\tilde{x}^{n_{1}}\tilde{y}^{n_{2}};$ $n_{1},n_{2}$ are integers.
The complete two-component wave function has to be defined as a combination
of two solutions $\psi _{n}^{\pm }$. In the initial frame the wave function
is%
\[
\Psi =\cos \frac{1}{2}\gamma \left( 
\begin{array}{c}
C^{+}\exp (-\frac{1}{2}i\Omega t-i\frac{E^{+}}{\hbar }t) \\ 
C^{-}\exp (\frac{1}{2}i\Omega t-i\frac{E^{-}}{\hbar }t)%
\end{array}%
\right) \exp (i\frac{p}{\hbar }z+D)+ 
\]%
\begin{equation}
\sin \frac{1}{2}\gamma \left( 
\begin{array}{c}
-C^{-}\exp (-\frac{1}{2}i\Omega t-i\frac{E^{-}}{\hbar }t) \\ 
C^{+}\exp (\frac{1}{2}i\Omega t-i\frac{E^{+}}{\hbar }t)%
\end{array}%
\right) \exp (i\frac{p}{\hbar }z+D),  \label{sol}
\end{equation}%
where $C^{+},C^{-}$ are constants. The wave function (\ref{sol}) \ describes
a state with the spin rotation.

Every $n$ is connected with $2(n+1)$ energy levels $E_{n}^{\pm }\equiv
E_{n}^{\sigma }$ with condition $E_{n}^{+}-E_{n}^{-}=-\hbar ^{2}\rho /m$.
The levels are determined from an algebraic equation of the power $n+1$. In
particular, first three energy levels are 
\begin{eqnarray}
E_{0}^{\sigma } &=&\frac{p^{2}}{2m}-\sigma _{3}\frac{\hbar ^{2}}{2m}\rho
+\epsilon _{0},  \label{E0} \\
E_{1}^{\sigma ,\tau } &=&\frac{p^{2}}{2m}-\sigma _{3}\frac{\hbar ^{2}}{2m}%
\rho \pm \tau +\epsilon _{1},  \label{E1} \\
E_{2}^{\sigma } &=&\frac{p^{2}}{2m}-\sigma _{3}\frac{\hbar ^{2}}{2m}\rho
+\epsilon _{2},  \label{E2} \\
E_{2}^{\sigma ,\tau } &=&\frac{p^{2}}{2m}-\sigma _{3}\frac{\hbar ^{2}}{2m}%
\rho \pm \tau +\epsilon _{2},  \label{E2d}
\end{eqnarray}%
\begin{eqnarray}
\epsilon _{n} &=&-\frac{\hbar ^{2}}{2m}%
[(n+1)(d_{11}+d_{22})+d_{1}^{2}+d_{2}^{2}],  \label{eps} \\
\tau &=&\frac{\hbar ^{2}}{2m}\sqrt{(d_{11}-d_{22})^{2}+d_{12}^{2}+f^{2}}.
\label{tau}
\end{eqnarray}%
The subscript of $E$ equals $n,$ energy levels are numbered by superscripts $%
\sigma ,\tau $. Two values of both superscripts $\sigma $ and $\tau $
corresponds two sign of $\sigma _{3}$ and two sign before $\tau $ in (\ref%
{E1}), (\ref{E2d}). Parameters $d_{kl},d_{k}$ are defined by equations%
\begin{eqnarray}
d_{11}^{2}+d_{12}^{2}-ifd_{12}-\text{\ }g_{1} &=&0,  \label{d11} \\
d_{22}^{2}+d_{12}^{2}+ifd_{12}-\text{\ }g_{2} &=&0,  \label{d22} \\
2d_{11}d_{12}+2d_{22}d_{12}-if(d_{22}-d_{11}) &=&0,  \label{d12} \\
2d_{11}d_{1}+2d_{12}d_{2}-ifd_{2} &=&0,  \label{d1} \\
2d_{22}d_{2}+2d_{12}d_{1}+ifd_{1}+b &=&0.  \label{d2}
\end{eqnarray}%
It may be shown with help these relations that if $f\rightarrow 0$ then $%
H\rightarrow 0$. In this limit solutions turn out into that of 2D axially
symmetric harmonic oscillator.

The wave function normalization%
\begin{equation}
\int \Psi ^{\ast }\Psi dxdy=1,  \label{norm}
\end{equation}%
where the integration is over all the cross-section, imposes one condition
on two coefficients $C^{+},C^{-}$ 
\begin{equation}
C^{+\ast }C^{+}+C^{-\ast }C^{-}=\frac{\sqrt{d_{11}d_{22}}}{\pi }\exp (\frac{%
d_{2}^{2}}{d_{22}}).  \label{nc}
\end{equation}%
The second one follows application scenarios. In particular, at the magnetic
resonance we have following expression for the average value of spin 
\begin{equation}
s_{3}=\frac{1}{2}\int \Psi ^{\ast }\sigma _{3}\Psi dxdy=\mp \frac{1}{2}\cos (%
\frac{2\mu H}{\hbar }t).  \label{s3}
\end{equation}%
The first condition of the magnetic resonance $C^{+}=\pm C^{-}$ is defined
by making the constant part of \ $s_{3}$ equal zero. It is achieved by the
initial polarization. The second condition $\gamma =\pi /2,$ is defined by
making the amplitude of the variable part equal maximum. It is achieved by
the adjustment of $\Omega $ or $H_{z}$. The same procedure for the magnetic
resonance definition is used in the next section for solutions of Dirac's
equation.

Desirable solutions does not exist if $d_{11}\geq 0,$ $d_{22}\geq 0,$ since
in this case the integral (\ref{norm}) tends to infinity and if 
\begin{equation}
4g_{1}\leq f^{2}\leq 4g_{2},  \label{gf}
\end{equation}%
since in this interval energy has complex values. At boundaries of (\ref{gf}%
) energy is real but at the lower and upper boundary $d_{11}=0$ and $%
d_{22}=0 $ respectively, i.e., solutions are not square integrable. With
help of (\ref{gf}) we may evaluate the forbidden zone for values of the
magnetic moment-or rather its $g$ factor-using the condition of the magnetic
resonance $\gamma =\pi /2$ or%
\begin{equation}
\hbar \Omega +\mu H_{z}=0,\text{ \ }\mu =g\frac{e\hbar }{2mc}  \label{mr}
\end{equation}%
(for electron $e=-|e|$ and $\Omega ,H_{z}$ has the same sign). From (\ref{mr}%
) and (\ref{ds3}) we obtain $f$ $=(1-g)eH_{z}/\hbar c.$ Then from (\ref{gf})
we find the forbidden zone for the $g$ factor%
\begin{equation}
1-\sqrt{1+\frac{4H^{2}}{H_{z}^{2}}}\leq g\leq 0,\text{ or }\ 2\leq g\leq 1+%
\sqrt{1+\frac{4H^{2}}{H_{z}^{2}}}.  \label{zone}
\end{equation}%
Usually $H\ll H_{z}$ and the zone represents a narrow vicinity of two points 
$g\approx 0$ and $g\approx 2$. The second inequality in Eq. (\ref{zone})
establishes an important result. Namely, the magnetic moment must be always
anomaluos.

Note that the Pauli principle \cite{paul} in the given case may be
reformulated as follows. Any state in a rotating magnetic field may have two
electrons with the opposite sense of the spin rotation.

In conclusion of this Section note that similar exact solutions are not
found for the Dirac equation. On the other hand solutions in the field of a
traveling circularly polarized electromagnetic wave are not found for the
Pauli equation but they exist for the Dirac equation. It is connected with
the unusual character of solutions. Solutions of the Dirac equation
considered below do not consist of a small and large two-component spinor.
Therefore they cannot be presented in the first approximation as solutions
of the Pauli equation

\section{The Dirac equation}

Consider exact solutions of the Dirac equation%
\begin{equation}
i\hbar \frac{\partial }{\partial t}\Psi +\mathbf{\alpha }(c\mathbf{\mathbf{p}%
}-e\mathbf{\mathbf{A}})\Psi +\beta mc^{2}\Psi =0  \label{Dir}
\end{equation}%
in the field of a powerful traveling circularly polarized wave and a
constant magnetic field directed along the axis of rotation (propagation).
Such a field corresponds to the potential $A_{x}=-\frac{1}{2}H_{z}y+\frac{1}{%
k}H\cos (\Omega t-kz),$ \ $A_{y}=\frac{1}{2}H_{z}x+\frac{1}{k}H\sin (\Omega
t-kz);$ $k=\varepsilon \Omega /c$\textit{\ }is the propagation constant,
where values $\varepsilon =+1$ and $\varepsilon =-1$ correspond to
propagation of fermions and wave in the same and opposite direction
respectively. Make manipulations analogously to previous section. However we
must take into account the coordinate $z$ in the translation of the wave
function $\Psi =\exp [-\alpha _{1}\alpha _{2}(\Omega t-kz)/2]\tilde{\Psi}$
as well as in the transformation of coordinates%
\begin{equation}
\tilde{\varphi}=\varphi -\Omega t+kz,\text{ \ }\tilde{r}=r,\text{ \ }\tilde{z%
}=z,\ \text{\ }\tilde{t}=t.  \label{D3}
\end{equation}

Further we consider the stationary case with the definite momentum $p:$ $%
\tilde{\Psi}=\exp (-iE\tilde{t}/\hbar +ip\tilde{z}/\hbar )\tilde{\Psi}_{s}.$
Then in coordinates $\tilde{x}=r\cos \tilde{\varphi},$ $\tilde{y}=r\sin 
\tilde{\varphi}$ exact solutions, somewhat more symmetric than that of the
Pauli equation, are $\tilde{\Psi}_{s}=\psi _{n}\exp D,$ where 
\begin{equation}
D=-\frac{1}{2}d(\tilde{x}^{2}+\tilde{y}^{2})+d_{1}\tilde{x}+d_{2}\tilde{y},
\label{psi}
\end{equation}%
$\psi _{n}$ is the spinor polynomial in $\tilde{x}$ and $\tilde{y}$. Such
solutions exist as for $eH_{z}<0,$ as for $eH_{z}>0$. In the first case $%
d=-eH_{z}/2\hbar c,$ the parameter $d$ is always positive since solutions
must decrease at infinity, $\ d_{1}=-id_{2},$

\begin{eqnarray}
d_{2} &=&\frac{mch\mathcal{E}_{0}}{2\hbar (\mathcal{E}-\mathcal{E}_{0})},%
\text{ \ }h=\frac{e}{kmc^{2}}H,  \label{dh} \\
\text{ \ }\mathcal{E} &=&\frac{E-cp\varepsilon }{mc^{2}},\text{ \ }\mathcal{E%
}_{0}\mathcal{=}\frac{2\hbar d}{\Omega m}.  \label{EE}
\end{eqnarray}%
As an example, we present here the normalized wave function of the ground
state for the first case

\begin{eqnarray}
\psi _{0} &=&N\sqrt{\frac{d}{2\pi }}\left( 
\begin{array}{c}
-\varepsilon h\mathcal{E} \\ 
(\mathcal{E}-1)(\mathcal{E}-\mathcal{E}_{0}) \\ 
h\mathcal{E} \\ 
-\varepsilon (\mathcal{E}+1)(\mathcal{E}-\mathcal{E}_{0})%
\end{array}%
\right) \exp D_{0},  \label{wf} \\
N &=&\frac{1}{\sqrt{(\mathcal{E}^{2}+1)(\mathcal{E}-\mathcal{E}%
_{0})^{2}+h^{2}\mathcal{E}^{2}}},  \label{N}
\end{eqnarray}%
where $D_{0}=(D-d_{2}^{2}/2d)$.

A distinguish feature of the solution (\ref{wf}) is that bispinor does not
consist of a small and large two-component spinor. It means that this
solution cannot be reduced to that of the Pauli equation. Solution in the
form of\emph{\ }an expansion in infinite series of terms like (\ref{sol}) is
possible. But such series evidently do not converge to a finite value.

Another distinguish feature is that energy of the ground state is defined by
an algebraic equation of the third order%
\begin{equation}
\mathcal{E}^{3}-(\mathcal{E}_{0}-\nu )\mathcal{E}^{2}-(1+\mathcal{E}_{0}\nu
+h^{2})\mathcal{E}+\mathcal{E}_{0}=0,  \label{eqE}
\end{equation}%
where $\nu =(2cp\varepsilon -\hbar \Omega )/mc^{2}.$ For $\nu ,h\rightarrow
0 $ Eq. (\ref{eqE}) has two positive and one negative roots $\mathcal{%
E\rightarrow }\pm 1,\mathcal{E\rightarrow E}_{0}>0$. Therefore a combination
of two solutions with positive roots may be used as a solution.

In the second case $d=eH_{z}/4\hbar c>0,$ $d_{1}=id_{2},$ 
\[
d_{2}=\frac{mch\mathcal{E}_{0}}{2\hbar (\mathcal{E}+\mathcal{E}_{0})}, 
\]%
and above expressions (\ref{wf}-\ref{N}) are still valid by substituting $%
\psi _{0}\rightarrow -\alpha _{1}\alpha _{3}\beta \psi _{0}.$

In the initial frame the wave function is 
\begin{equation}
\Psi =\exp [-iE\tilde{t}/\hbar +ip\tilde{z}/\hbar -i\alpha _{1}\alpha
_{2}(\Omega t-kz)/2+D]\psi _{n}.  \label{wfd}
\end{equation}%
According to the procedure of the previous section consider the average
value of spin 
\begin{equation}
s_{3}=-\frac{i}{2}\int \Psi ^{\ast }\alpha _{1}\alpha _{2}\Psi dxdy
\label{sd}
\end{equation}%
and its temporal evolution. The wave function consists of two solutions with
two positive roots of Eq. (\ref{eqE}): $\Psi =C_{1}\Psi _{1}+C_{2}\Psi _{2}.$
The normalization integral (\ref{norm}) and the equality to zero of \ the
constant part of $s_{3}$ results in following values of normalization
constants $C_{1}=\cos \theta ,$ \ $C_{2}=\sin \theta ,$%
\begin{equation}
\cos 2\theta =\frac{h^{2}\Pi ^{2}\mathcal{E}^{-}}{(\Pi +1)^{2}[(\mathcal{E}%
_{0}{}^{2}-\Pi ^{2})\mathcal{E}^{+}+2\Pi (\Pi -1)\mathcal{E}_{0}]},
\label{teta}
\end{equation}%
where 
\begin{equation}
\Pi =\mathcal{E}_{1}\mathcal{E}_{2},\text{ \ }\mathcal{E}^{+}=\mathcal{E}%
_{1}+\mathcal{E}_{2},\text{ \ }\mathcal{E}^{-}=\mathcal{E}_{1}-\mathcal{E}%
_{2}.  \label{E12}
\end{equation}%
For the average spin we obtain 
\begin{equation}
s_{3}=h^{2}\Pi N_{1}N_{2}\exp \left[ -\frac{(d_{2}^{\prime }-d_{2}^{\prime
\prime })^{2}}{d}\right] \sin 2\theta \cos \Phi ,  \label{s}
\end{equation}%
where $N_{1},N_{2}$ and $d_{2}^{\prime },d_{2}^{\prime \prime }$ is the
normalization parameter (\ref{N}) and the parameter (\ref{dh}) for $\mathcal{%
E}_{1},\mathcal{E}_{2}$ respectively, $\Phi $ defines the frequency of the
spin oscillation%
\begin{equation}
\omega =\frac{\Phi }{t}\equiv \frac{\mathcal{E}^{-}}{\hbar }.  \label{ft}
\end{equation}

It may be shown that for small $\eta ,v,h$ in the first approximation the
amplitude maximum of $s_{3}$ is realized at $\mathcal{E}_{0}=1$. It is
easily to see that this equality is the condition of the magnetic resonance (%
\ref{mr}) for $g=2$. However next approximations lead to conflicting
results. Once again emphasize the solution (\ref{sol}) is not an
approximation of (\ref{wf}). The solution (\ref{wf}) represents a new class
of solutions.

In the general case the amplitude of the average spin as well as the
frequency of the spin oscillation is a complicated function of all
parameters. However both variables can be calculated in accordance with (\ref%
{s}), (\ref{ft}), adjusted and tested experimentally.

\section{Appendix. The transformation for point rotation coordinate frames}

The concept of the point rotation frame arises in electrooptics, when light
propagates through electrooptical crystal with the rotating optical
indicatrix (index ellipsoid) \cite{sm}. The optical indicatrix is an example
of such a frame however we attach a more fundamental meaning to this
concept. The concept is applicable to any rotating field, in particular to
rotating magnetic field in quantum mechanics. A distinguish feature of the
point rotation frame is the existence of the rotation axis at every point.
All points in the plane perpendicular to the rotation axis are equivalent
but time in frames rotating at different frequencies may be different.
Centrifugal forces are absent in this frame. Coordinates of the frame are an
angle and time; the frequency of rotation is a parameter \cite{job}. Such
non-rotating frames are used in quantum field theory for a long time.

Above it is shown that the first step in studying solutions of the Pauli and
Dirac equation in the rotating electromagnetic field is the translation of
the wave function into the rotating frame. This translation must accompanied
by the transformation of coordinates and time. For finding of above
solutions the Galilean transformation (\ref{D2}) or (\ref{D3}) is used.
However there is possible a phase mismatch if the real transformation is
non-Galilean and time in the initial and rotating frame is different. The
problem in a sense is similar to that in mechanics where the consideration
in the system of the mass center is preferable. The correct result in
mechanics can be obtained only if the Lorentz transformation is used for
the\ transition into this system. In the opposite case some corrections of
the result are necessary. In this sense the frame with the resting magnetic
field is of particular importance as an analog of the mass center in
mechanics.

The general linear 2D transformation for the point rotating frames was
considered in \cite{job}. It was assumed that symmetry exists by the
exchange angle $\leftrightarrow $ time and shown that in this condition a
non-trivial transformation together with the Lorentz transformation is
possible. This transformation possesses unusual properties and enables the
existence of allowable frequency regions with lower and upper boundaries
(see an example in \cite{arx}). From the general physical viewpoint it seems
more preferable in comparison with the Galilean (with infinite frequencies)
or Lorentz (with the sole limiting frequency) transformation. However the
transformation cannot be uniquely determined with help of the assumption on
the symmetry only. In \cite{job} the 2D transformation was used for a
phenomenological description of a circularly polarized light wave in the
electrooptical single-sideband modulator \cite{pat}, \cite{jpc}. This
description is similar to an elastic collision in mechanics, however such a
consideration is not applicable to quantum mechanics.

The 3D Galilean transformation (\ref{D3}) is used for finding solutions of
the Dirac equation in the field of traveling circularly polarized wave. Some
aspects of the general 3D transformation is considered in \cite{arx}. This
transformation also cannot be uniquely determined. In both cases there is
lacking a principle which would allow determining the transformation
explicitly.

\section{Conclusion}

We have presented the new class of exact solutions of the Pauli and Dirac
equation in rotating electromagnetic fields. Their properties differ
significantly from known presently exact solutions. That especially is
related to the Dirac equation. In the standard approach the Pauli equation
is not the first approximation of Dirac's equation for such solutions. The
obtained solutions may be a basis for experiments at large values of
parameters, in particular, for relativistic electron.

We have discussed non-Galilean transformation. Main problem is the absence
of a principle for determining the transformation. Main advantage of such a
transformation is the possibility of different lower and upper frequency
boundaries for different fields. This is important from the general physical
viewpoint and calls for detail investigations of this issue.

\end{document}